# AN EMPIRICAL STUDY OF SOFTWARE REUSE BY EXPERTS IN OBJECT-ORIENTED DESIGN [1]

*Jean-Marie BURKHARDT*Δ *and Françoise DETIENNE**

\* INSTITUT NATIONAL DE RECHERCHE EN INFORMATIQUE ET EN AUTOMATIQUE (INRIA)
Projet de Psychologie Ergonomique
Domaine de Voluceau, Rocquencourt, B.P. 105  78 153 Le Chesnay Cedex France
E-mail: Jean-Marie.Burkhardt@inria.fr , Francoise.Detienne@inria.fr

Δ UNIVERSITE PARIS V RENE DESCARTES
Laboratoire d'Ergonomie Informatique
45 rue des Saint -Pères 75 270 Paris cedex 06 France

**ABSTRACT :** This paper presents an empirical study of the software reuse activity by expert designers in the context of object-oriented design. Our study focuses on the three following aspects of reuse : (1) the interaction between some design processes, e.g. constructing a problem representation, searching for and evaluating solutions, and reuse processes, i.e. retrieving and using previous solutions, (2) the mental processes involved in reuse, e.g. example-based retrieval or bottom-up versus top-down expanding of the solution, and (3) the mental representations constructed throughout the reuse activity, e.g. dynamic versus static representations. Some implications of these results for the specification of software reuse support environments are discussed.

## 1 FRAMEWORK AND GOALS

Software Reuse is currently one of the most active and creative research areas in Computer Science. This is mainly because software quality and productivity are assumed to be greatly increased by maximising the (re)use of (part of) prior design products instead of repeatedly designing from scratch (Krueger, 1989). Currently, the predicted level of reuse has not yet been reached, due to technical, organisational and ergonomic factors (Tracz,1987). This paper presents an empirical study[1] of the software reuse activity. The originality of this study is twofold. First, it is one of the few experimental investigations into the software reuse activity by expert designers. Second, it examines reuse in the context of object-oriented programming which is assumed by its advocates to facilitate reuse. Few studies (Détienne, 1991; Lange & Moher, 1989) have analysed reuse in that context.

Our study focuses on the three following aspects of reuse : the interaction between design and reuse processes, the mental processes involved in reuse, and the mental representations constructed throughout the reuse activity.

First, we analyse how some **design** processes, e.g. constructing a problem representation, searching for and evaluating the solution(s), and **reuse** processes, i.e. retrieving and using previous solution(s), may interact. For example, recalling solutions may lead to a revision of the currently-developed solution and retrieving a past solution may produce the addition of constraints to the representation of the current design problem.

Second, we analyse some of the mental processes involved in the reuse activity. We investigate the retrieval processes involved in reuse. One issue is whether retrieval may be based on contextual and episodic cues, as shown in studies on the elicitation of category members (Walker & Kintsch, 1985) , or on abstract features, mainly used by the software community as a basis for information retrieval. We also investigate the direction of the solution expansion on which the reuse activity is based, i.e. bottom-up versus top-down.

---

[1] This research is partially sponsored by the European Esprit III SCALE 6334 project (System Composition and Large Grain Component Reuse Support)

**Figure 1:** Successive tasks in the experiment

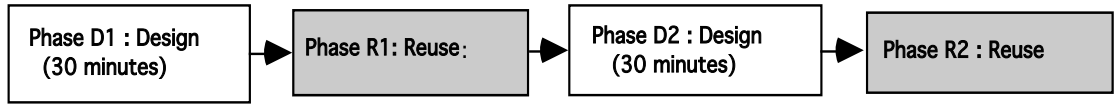

Third, we are interested in the characteristics of the mental representations constructed during reuse. Psychological studies (Adelson & Soloway, 1988) pointed out that experts involved in non-routine design often simulated their more or less complete solution. The distinction is made between "dynamic" mental representations, i.e. involving simulating and enacting of a mental model, and "static" mental representations, i.e. involving static statements such as states and properties. We assume that designers may favour one such representation rather than another depending on the task involved (i.e. reuse versus design).

## 2 METHOD

### 2.1 Subjects
Seven experts in Object-Oriented Programming participated in this experiment. They had between three and five years' experience with C++ through real projects and they knew at least five different programming languages (C++, C, Pascal, Lisp and Prolog). All belonged to different teams at INRIA. They were familiar with the management domain.

### 2.2 Design problem
The problem to be solved, a management problem, involved designing a system to compute manufacturing capacity from existing spare stock and a list of required components to build a vehicle. This problem appeared to be very suitable for an object-oriented context of design.

### 2.3 Experimental design and procedure
The experimental design consisted of alternating between two "design phases", i.e. analysing the problem and developing a solution, and two "reuse phases", i.e. describing elements which the designer would like to "get from an intelligent library" in order to reuse them (cf. figure 1). The actual activity of the designers implies numerous shifts between developing a solution and retrieving previous solutions. Separating these phases in our experimental setting, although quite artificial, should allow us to examine the interaction between some design processes, e.g. understanding the problem and evaluating the solution(s), and reuse processes.

The two design phases had a prefixed 30-minute duration that was sufficient to develop an advanced solution before being interrupted by the experimentor to introduce the reuse phase. The subjects were told that they were not supposed to implement their solution during the experiment. Nevertheless, they were encouraged to produce a sufficiently detailed solution which a C++ programmer could easily implement later on.

Each reuse phase consisted of asking the designers what they would wish to get from the library at the current state of their solution. We pointed out that the content of the library, in which the designers were supposed to find the to-be-reused elements[2], was not restricted to the usual software components. It was emphasised that the designers could obtain whatever they needed, including for example, any of their past designs, documentation etc. It was made clear that a complete software solution to the problem at hand could not be reused.

The reuse phases had no fixed duration, since they stopped when the designers no longer required other to-be-reused elements. Then, they were encouraged to go on with the design, and they were told that, unfortunately, no element they had just asked for was available.

Sessions were gone through individually. The subject was given a pen and a set of numbered sheets of paper. S/he was asked to "think aloud".

### 2.4 Collected data and method of analysis
Collected data included tape-recorded verbal protocols, notes or graphics, and the successive versions of each subject's solution.

We identified the elements of the solutions developed in each phase and the to-be-reused elements mentioned. Then, we characterised how they were represented, e.g. dynamic versus static representations, and example-based versus formal attributes. Furthermore, we made a mapping between the to-be-reused elements and the elements of the current solutions. Finally we investigated some of the mental processes involved in the task by analysing verbal protocols. A complete description of our analysis method may be found in (Détienne, Rouet, Burkhardt, Chatel & Deleuze-Dordron, 1994).

---

[2] We will refer to the "expected elements to be available for reuse" as the "to-be-reused elements".

## 3 RESULTS

First, we should point out that there was a great variability between the to-be-reused elements described for the same problem even if the designers had the same kind of expertise in the problem and in the programming domains. There was also a large variability of descriptors used for the same to-be-reused elements. These elements were defined at various levels of granularity.

This section is organised according to three main lines: interactions between design and reuse activities, mental processes involved in reuse, and features of the mental representations constructed in reuse.

### 3.1 Interaction(s) between Design and Reuse

**Looking for a model versus plug-in modules** Our data showed that the to-be-reused elements could have various statuses depending on how the designers considered them. In some cases, the designers were explicitly looking for a model, from the problem domain and/or the software solution domain, on which to base their design activity. In other cases, the designers were looking for software units that they could plug into their current solution, with a greater or smaller amount of modification. It appeared that one particular element might be considered as a model at one moment and as a plug-in component at another time. Our findings are consistent with results of previous studies on software design (Lange & Moher, 1989) or software enhancement by reusing supplied classes (Rosson & Carroll, 1993), and illustrate that models can also come from the current or other analogical problem domains.

**Trade-off between design and reuse costs** We observed that the designers evaluated the relative costs of reuse versus design, in order to decide whether or not to reuse a component. They made explicit in their verbalisation, the trade-off between design and reuse costs, as summarised in figure 2.

**Figure 2**: Trade-off between design and reuse costs

|  | Reuse activity | |
|---|---|---|
| **Design activity** | low cost | high cost |
| low cost | => Design | => Design |
| high cost | => Reuse | |

Although this trade-off seems to be spontaneously associated to the process of design with reuse, great difficulty in evaluating such costs had been reported in software abstract data type reusability assessment by untrained reusers (Woodfield, Embley, & Scott, 1987) : the subjects tended to be mostly influenced by irrelevant factors, e.g. size of the abstract data type, and failed to consider relevant aspects, e.g. the time needed to modify the abstract data type.

**Evaluation/revision of the solution** We observed that example-based descriptors of the to-be-reused element often allowed the designers to mention validity constraints related to similar problem contexts, e.g. "an associative table with string keys is not the most general in theory, but in practice it's enough to solve any prolem, especially awk only has strings of characters". This kind of information was used to evaluate whether or not the to-be-reused element was valid for the solution context at hand. Judging validity may lead to solution revision as was observed once.

**Addition of design constraints** We observed on several occasions that referring to to-be-reused elements could add new constraints to the problem representation. Such an effect was also reported by De Vries (1993) in the field of architectural design. She showed that the provision of a large number of examples that were organized in three abstract-to-concrete levels, had two implications on the problem representation that the designers built : new constraints were added to the problem representation, and the level at which constraints were initially considered became significantly higher.

### 3.2 Mental Processes involved in Reuse

**Example versus formal attribute based retrieval process** On the basis of the assumption that retrieval cues are more example-based than abstract-based, we analysed the distribution of the example-based versus formal-based descriptors associated to the to-be-reused elements. We consider that the experts referred to an example whenever they made explicit references to concrete/existing elements or to their personal experience. Moreover, we distinguished whether the examples were explicitly referred to via personal experience or not.

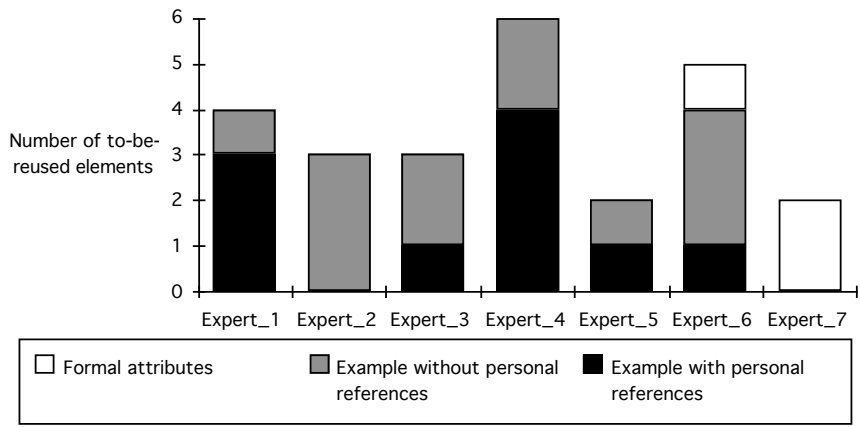

**Figure 3** : Number of to-be-reused elements referred to via example versus formal attribute.

Figure 3 presents the number of to-be-reused elements referred to via an example with or without personal references and referred to via formal attributes. First, it shows that the described to-be-reused elements were mainly associated to example(s) whether or not personal references were evoked. It must be pointed out that the use of examples may be explained by an attempt to be more understandable, using more imagery for the sake of the experimenter. Furthermore, it permits the designers to quickly and more globally express their ideas. Finally as noted by Kruger (1993), it is difficult to decide whether examples were actually used to generate the solution, or were only used to interpret and verbalize it.

Second, 45.4% (10/22) of the examples were associated to personal references. This result reveals the use of episodic knowledge in reuse. It is consistent with the findings of Walker and Kintsch (1985) in an elicitation of category members task also involving knowledge retrieval and elicitation.

**Top-down and bottom-up processes** Figure 4 shows the total number of the to-be-reused elements that were mentioned (all subjects are combined) during the reuse phases, depending on the relationship with the elements that were previously developed during the design phase. The first row gives the absolute number of to-be-reused elements, described during each reuse phase, and that were linked to at least two distinct entities of the current developed solution. The second row, called addition, presents aspects that were not previously envisaged in the design phase. The observed decrease in the R2 phase is explained by the fact that designers included, during the second phase of design (D2),

**Figure 4**: Relationships between to-be-reused elements and solution elements

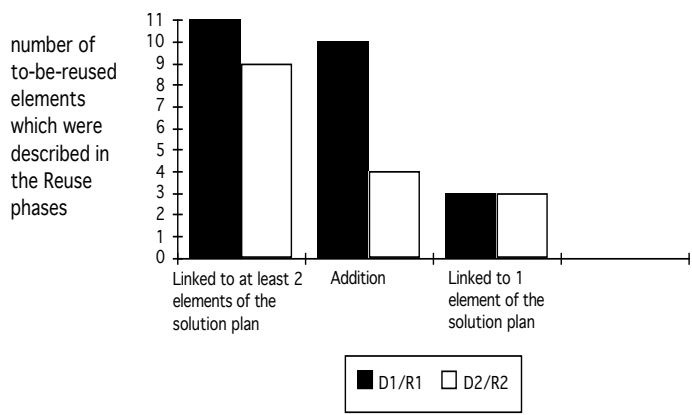

some of the previously elicited aspects (e.g. interface, storage management) during R1.

The first and second categories appear to be mainly represented in our collected data. We could interpret these results as being two distinct ways of expanding the current solution. On the one hand, the experts adopted a bottom-up approach : they tended to link the previously defined entities through a new (to-be-reused) entity which was more abstract. On the other hand, they adopted a top-down approach since they tended to expand their current solution by adding new (to-be-reused) entities.

### 3.3 Mental Representations constructed in Reuse

**Solution-mediated description of to-be-reused elements** We distinguished between two ways of describing the to-be-reused elements. The first way is called "direct description": functions or properties of to-be-reused elements are simply described (e.g. "a class called *Application* that manages the interface ... there already are standard menus"). The second way is called "solution-mediated description": the to-be-reused elements are mainly described through their instantiation in the current solution (e.g. "If the *Model* class is like ... that is, whenever I want to create an instance ... let's say *M1*, then as it inherits from the persistent objects class ... *M1* is automatically popped into the list"). Our results show that the to-be-reused elements were mainly associated to descriptions instantiated in the current solution (22 out of 25 observations, all protocols being combined). This behaviour does not appear to depend on the relationship between the solution and the to-be reused element. This phenomenon may be due to the process of progressively going deeper into the matching between the source and the target situation.

**Mental representation in the Design phase versus in the Reuse phase** We distinguished between "dynamic" mental representations, i.e. involving simulating and enacting a mental model (e.g. control flow based representations), and "static" mental representations, i.e. involving static statements like states and properties (e.g. object-based representations). The distinction was investigated here by analysing subject verbalisations. A first result is that there were more static representations than dynamic representations whatever the phase. Figure 5 compares the percentage of dynamic representations in the first phases of design (D1) and Reuse (R1) for each subject. It shows that the experts tended to use more dynamic representation in D1 than R1. Similar results are found when comparing the second phases of design (D2) and Reuse (R2). Nevertheless, statistical tests do not reach a significant level (Wilcoxon matched-pairs Signed-ranks).

### 4. LIMITATIONS / IMPLICATIONS OF THE STUDY

First, we must point out several limitations of our study. The low number of subjects makes it difficult to generalise our results. Our experimental procedure separating design from reuse was quite artificial, and did not allow us to collect data on the process of source comprehension, or on the processes of modifying and integrating reusable

**Figure 5:** Percentage of dynamic representations in the first phases of design (D1) and Reuse (R1)

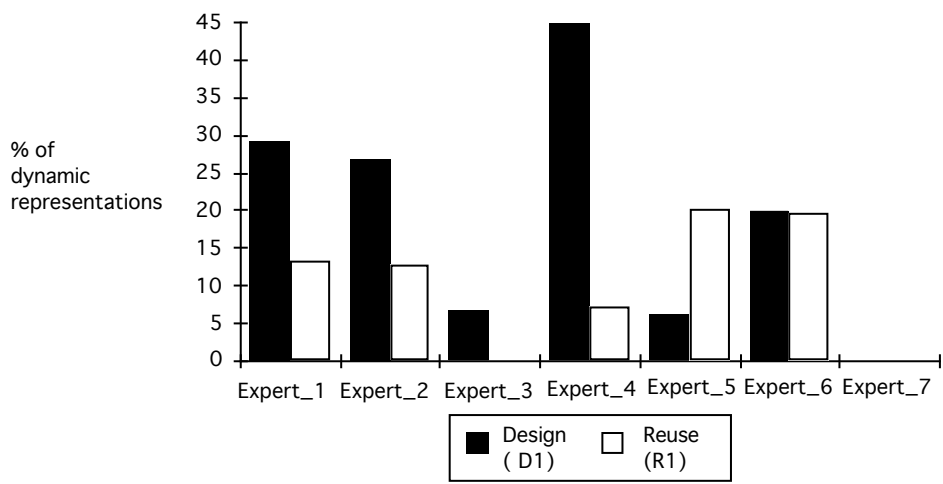

components. We are currently examining these points in another experiment.

Several results are particularly important for the specification of a software reuse support environment. We observed that designers often evoked the to-be-reused elements through examples. An interesting result is that the use of an example allows the subjects to evoke validity constraints related to similar problem contexts. Thus examples seem important in the retrieval process as well as in the selection process of a reusable component. This contradicts the idea of the software community that favours abstract and formal ways of displaying and organizing information for reuse.

A dynamic view of the solution was observed to be used more often in the design phase than in the reuse phase. However, such a view may also be important when designers integrate reusable elements in their target solution, in particular, for evaluating completeness and correctness. More generally, an environment supporting design with reuse should allow the designers to have either a dynamic view or a static view of their solution depending on the process they are involved in.

Finally, we showed that either a top-down approach or a bottom-up approach may be involved in the reuse activity. Whereas the former is often supported by tools, the latter is considered to be contrary to hierarchical prescribed methods. Such an approach could also be supported in as much as it represents a natural way of designing with reuse.

**ACKNOWLEDGEMENTS**

Special thanks to Willemien Visser for her comments on an earlier draft of this paper.